\documentclass[aps,prd]{revtex4}
\usepackage{subfigure,graphics} 
\usepackage{flafter} 
\begin{document} 

\title{
Density fluctuations from warm inflation} 
\author{Chris Graham}
\author{Ian G. Moss}
\email{ian.moss@ncl.ac.uk}
\affiliation{School of Mathematics and Statistics, Newcastle University, NE1 7RU, UK}

\date{\today}


\begin{abstract}
Thermal fluctuations provide the main source of large scale density perturbations in warm
inflationary models of the early universe. For the first time, general results are obtained for the
power spectrum in the case when the friction coefficient in the inflaton equation of motion depends
on temperature. A large increase in the amplitude of perturbations occurs when the friction
coefficient increases with temperature. This has to be taken
into account when constructing models of warm inflation. New results are also given for the thermal
fluctuations in the weak regime of warm inflation when the friction coefficient is relatively
small.
\end{abstract}
\pacs{PACS number(s): }

\maketitle
\section{introduction}

Observations of the cosmic microwave background have provided strong evidence that
the primordial density fluctuations have a power spectrum very close to that predicted
by inflationary models of the very early universe \cite{guth81,linde82,albrecht82}. 
According to the standard theory,  these density fluctuations can arise from quantum vacuum
fluctuations during inflation \cite{liddle}. Another possibility, which also fits the data, 
is that  thermal fluctuations provide a source
of density fluctuations \cite{moss85,bererafang95}.  This is what happens in the warm inflationary
scenario \cite{Berera:2008ar}.

During an inflationary era particles are being produced continually but their density is also
rapidly diminished by the expansion of the universe. What distinguishes warm inflationary models is
that this particle production is sufficiently strong, compared to the effects of expansion, to
produce a non-negligible particle density. 
Warm inflation occurs when the quantum statistical fluctuations in the particle number are large enough to
influence the classical inflaton field and produce density fluctuations \cite{berera00,Hall:2003zp}.

The description of coupled fluctuations between classical sources and quantum radiation fields can be
traced back to Einstein's classic discussion of a mirror immersed in black body radiation
\cite{einstein}.
Stochastic fluctuations in the radiation field lead to fluctuations of the mirror.
In an analogous way, fluctuations in the radiation field are transfered into inhomogeneities in the inflaton field.
In a non-expanding universe, standard thermal field theory methods can be used to reduce
the description of the system to a pair of coupled stochastic equations, one for the inflaton
and one for the radiation density \cite{gleiser94,berera98}. 
The simplest case to analyse is where the radiation
is close the thermal equilibrium, and then the particle production rate, the dissipation 
in the inflaton equation of motion and the source terms in the inflaton equation are all related by
a single friction coefficient $\Gamma$ \cite{gleiser94,berera98,Berera:2007qm,Graham:2008vu}. 
 
There has been progress in extending the theory of thermal fluctuations to an expanding 
universe \cite{Berera:2004kc}, but it seems that the basic equations 
which we require can be determined
from the equivalence principle \cite{Hall:2003zp,Moss:2007cv}. 
This approach will be pursued further in Sect.
III, where we present some new modifications to stochastic source terms in curved space. Our
results, however, are broadly in line with those presented elsewhere which imply a close to
scale-free power spectrum of density perturbations with small corrections depending on the
slow-roll parameters \cite{taylor00,oliveira02,Hall:2007qw}.

The main objective in this paper is to present results for the power spectrum of fluctuations
in inflationary models where the friction coefficient $\Gamma$ depends on the temperature $T$. 
Temperature dependence seems to be a common feature in explicit calculations of the 
particle production rates. In particular, there are high-temperature models of warm inflation with
friction coefficient $\Gamma\propto T^{-1}$ \cite{hosoya84,Moss:2006gt} and the two-stage decay
models
\cite{berera03} with friction coefficient $\Gamma\propto T^3$ \cite{Moss:2006gt}. High-temperature
models are difficult to realise without producing large thermal corrections to the potential, which
make the models unstable \cite{Moss:2008yb}, and most consistent warm inflationary scenarios use
the two-stage decay mechanism (see Ref. \cite{Berera:2008ar} for a review).

Individual models with temperature dependent friction coefficients have been 
examined using the  numerical code written for Ref. \cite{Hall:2003zp}, 
which solves coupled perturbation 
equations for the time evolution of the inflaton, radiation and metric. The model 
with $\Gamma\propto T^{-1}$ described in Ref. \cite{Hall:2003zp} produced an 
oscillatory power spectrum, 
but it has not been possible previously to formulate general laws for the functional dependence of
the power spectrum.

An analysis of the coupled perturbation equations is presented in Sect. IV. We have been able to
find approximate solutions to the perturbation equations
on sub-horizon scales to determine general formulae for the power spectrum. 
These formulae have been checked against a numerical solution of the stochastic perturbation
equations in section V.

The most important new results apply to the `strong' regime of warm inflation where 
$\Gamma\gg 3H$ and $H$ is the expansion rate. There is a large enhancement of the density
perturbation amplitude 
when $\Gamma\propto T^3$, and a large diminution when
$\Gamma\propto T^{-1}$. In practice, the power spectrum of density fluctuations is fixed by
observation, and so the enhancement of the amplitude is equivalent to a reduction of the
inflationary energy scale, and visa versa. We have also found that the oscillations of the power
spectrum found in Ref.  \cite{Hall:2003zp} are removed by the ensemble averaging.
These results are an essential ingredient for the construction of warm inflationary models.

\section{warm inflation}

We shall assume that particles are produced during the inflationary era and that
the particle interactions are strong enough to produce a thermal gas of
radiation with temperature $T$. The necessary conditions for this to happen where discussed in Ref. \cite{Graham:2008vu}. Warm inflation is said to occur when $T$ is
larger than the energy scale set by the expansion rate $H$. The particle production is associated
with a dissipative effect on the inflaton, whose equation
of motion becomes \cite{moss85,berera95}
\begin{equation}
\ddot\phi+(3H+\Gamma)\dot\phi+V_\phi=0
\end{equation}
where $\Gamma(\phi,T)$ is a friction coefficient, $H$ is the Hubble parameter
and $V_\phi$ is the $\phi$ derivative of the inflaton effective potential $V(\phi,T)$.
The effectiveness of warm inflation can be parameterised by a parameter $r$,
defined by
\begin{equation}
r={\Gamma\over 3H}
\end{equation}
When $r\gg 1$ the warm inflation is described as being in the strong regime and when $r\ll1$ the
warm
inflation is in the weak regime.

Consistent models of warm inflation \cite{Moss:2008yb} require a suppression of thermal corrections
to the inflaton potential, so that the effective potential separates into inflaton and radiation
components
\begin{equation}
V(\phi,T)=V(\phi)+\rho_r(T),
\end{equation}
where $\rho_r$ is the radiation density
\begin{equation}
\rho_r={\pi^2\over 30}g_*T^4.
\end{equation}  
In this case, the time evolution is described by the equations
\begin{eqnarray}
&&\ddot\phi+(3H+\Gamma)\dot\phi+V_\phi=0,\\
&&\dot\rho_r+4H\rho_r=\Gamma\dot\phi^2,\\
&&3H^2=4\pi G\left(2V+2\rho_r+\dot\phi^2\right)
\end{eqnarray}
During inflation we apply a slow-roll approximation and drop the highest
derivative terms in the equations of motion,
\begin{eqnarray}
&&3H(1+r)\dot\phi+V_\phi=0,\label{sr1}\\
&&4H\rho_r=\Gamma\dot\phi^2,\label{sr2}\\
&&3H^2=8\pi G V\label{sr3}
\end{eqnarray}
The validity of the slow-roll approximation depends on the slow roll parameters
defined in \cite{Hall:2003zp},
\begin{equation}
\epsilon={1\over 16\pi G}\left({V_\phi\over V}\right)^2,\qquad
\eta={1\over 8\pi G}\left({V_{\phi\phi}\over V}\right),\qquad
\beta={1\over 8\pi G}\left({\Gamma_\phi V_\phi\over \Gamma V}\right)
\label{slowrp}
\end{equation}
The slow-roll approximation holds when $\epsilon\ll 1+r$, $\eta\ll 1+r$ and
$\beta\ll 1+r$. Any quantity of order $\epsilon/(1+r)$ will be described as being
first order in the slow roll approximation.

\section{Inflaton fluctuations}

Thermal fluctuations are the main source of density perturbations in warm
inflation. Thermal noise is transfered to the inflaton field mostly on small
scales. As the comoving wavelength of a perturbation expands, the thermal
effects decrease until the fluctuation amplitude freezes out \cite{berera00}. This may occur
when the wavelength of the fluctuation is still small in comparison with
cosmological scales.

In this section we shall review the equation for the inflaton fluctuations in the case where the
friction coefficient is independent of temperature, i.e.
\begin{equation}
\Gamma=\Gamma(\phi).
\end{equation}
We shall see how the stochastic inflaton dynamics transfers to an expanding universe and obtain new
results for the case $\Gamma\ll 3H$. 

The behaviour of a scalar field interacting with radiation can be analysed
using the Schwinger-Keldysh approach to non-equilibrium field theory 
\cite{schwinger61,keldysh64}. In flat spacetime, when the small-scale behaviour of the 
fields is averaged out, a simple picture emerges in which the field can
be described by a stochastic system whose evolution is determined by a Langevin
equation \cite{calzetta88}. This takes the form
\begin{equation}
-\nabla^2\phi(x,t)+\Gamma\dot\phi(x,t)+V_\phi=(2\Gamma T)^{1/2}\xi(x,t),\label{stoc}
\end{equation}
where $\nabla^2$ is the flat spacetime Laplacian and $\xi$ is a stochastic
source. For a weakly interacting radiation gas the probability distribution of the source term can
be approximated by a localised gaussian distribution with correlation function 
\cite{gleiser94,Berera:2007qm},
\begin{equation}
\langle\xi(x,t)\xi(x',t')\rangle=\delta^{(3)}(x-x')\delta(t-t').
\end{equation}
We shall restrict ourselves to this gaussian noise approximation.

We can use the equivalence principle to adapt the flat spacetime Langevin
equation to an expanding universe during a period of warm inflation when $T>H$. The Langevin
equation will retain its local form as long as the microphysical and thermal scales in the problem
are small compared to the cosmological scale \cite{Berera:2007qm,Berera:2008ar}. However, the rest
frame of the fluid will have a non-zero $3-$velocity $v_r^\alpha$ with respect to the
cosmological frame and we must include an advection term. The Langevin 
equation becomes \cite{Moss:2007cv}
\begin{equation}
\ddot\phi(x,t)+(3H+\Gamma)\dot\phi(x,t)+
\Gamma a^{-2}v_r^\alpha\partial_\alpha\phi(x,t)
+V_\phi-a^{-2}\partial^2\phi(x,t)=(2\Gamma_{\rm eff} T)^{1/2}\xi(x,t)\label{langa}
\end{equation}
where $a$ is the scale factor and $\partial^2$ is the Laplacian in an expanding frame with
coordinates $x^\alpha$. The correlation function for the noise expressed in terms of the 
comoving cosmological coordinates becomes,
\begin{equation}
\langle\xi(x,t)\xi(x',t')\rangle= a^{-3}(2\pi)^2
\delta^{(3)}(x-x')\delta(t-t').
\end{equation}
A new parameter $\Gamma_{\rm eff}$ has been introduced because at this stage we 
cannot determine the effects of the expansion on the noise term.

Next we treat the source term as a small perturbation and expand the inflaton
field
\begin{equation}
\phi(x,t)\to\phi(t)+\delta\phi(x,t)
\label{ipe}
\end{equation}
where $\delta\phi$ is the linear response due to the source $\xi$.
This expansion is substituted into the Langevin equation and then we take the 
Fourier transform,
\begin{equation}
\delta\ddot\phi({\bf k},t)+(3H+\Gamma)\delta\dot\phi({\bf k},t)+V_{\phi\phi}\delta\phi({\bf k},t)
+k^2a^{-2}\delta\phi({\bf k},t)=(2\Gamma_{\rm eff} T)^{1/2}\xi({\bf k},t).\label{langa}
\end{equation}
The inflaton will also generate metric inhomogeneities, but with a suitable choice of
gauge, these can be discarded on sub-horizon scales (see \cite{Moss:2007cv}). 
We shall use a uniform expansion rate gauge. Eq. (\ref{langa}) applies on
scales which are intermediate between the thermal averaging scale and the horizon scale.
Later, we use a matching argument to extend the fluctuations to large scales.

The analysis of the Langevin
equation can be simplified by introducing a new
time coordinate $z=k/(aH)$ and using the slow roll parameters 
(\ref{slowrp}). We are
led to the equation for $\delta\phi({\bf k},z)$,
\begin{equation}
(1-\tilde\epsilon)^2\delta\phi^{\prime\prime}-(3r+2)(1-\tilde\epsilon)z^{-1}\delta\phi'
+\tilde\epsilon'\delta\phi'+3\eta z^{-2}\delta\phi+\delta\phi
=(2\Gamma_{\rm eff} T)^{1/2}(1-\tilde\epsilon)^{1/2}\hat\xi
\end{equation}
where a prime denotes a derivative with respect to $z$ and $\tilde\epsilon=\epsilon/(1+r)$. This
reduces to
\begin{equation}
\delta\phi^{\prime\prime}-(3r+2)z^{-1}\delta\phi'
+\delta\phi
=(2\Gamma_{\rm eff} T)^{1/2}\hat\xi
\end{equation}
when we keep only the leading terms in the slow roll approximation. The noise term has been rescaled
so that the correlation function is now
\begin{equation}
\langle\hat\xi({\bf k},z)\hat\xi({\bf k}',z')\rangle=(2\pi)^3
\delta^{(3)}({\bf k}-{\bf k}')\delta(z-z')k^{-3}.\label{cf}
\end{equation}

The perturbation equations can be solved using Green function techniques. The
solution is
\begin{equation}
\delta\phi=\int_z^\infty G(z,z')(z')^{1-2\nu}\,
(2\Gamma_{\rm eff}T)^{1/2}\xi(z')\,dz'.
\label{usol}
\end{equation}
The retarded Green function $G(z,z')$ can be found in terms of Bessel functions,
\begin{equation}
G(z,z')={\pi\over 2}z^{\nu} z^{\prime\nu}
\left(J_{\nu}(z)Y_{\nu}(z')-J_{\nu}(z')Y_{\nu}(z)\right).
\quad\hbox{ for }z<z'.
\label{rgf}
\end{equation}
where
\begin{equation}
\nu=\frac32(1+r)={\Gamma+3H\over 2H}.\label{defnu}
\end{equation}
Corrections due to the time dependence of $\nu$ are similar in size to terms
which we have already discarded in the slow roll approximation.

\subsection{Inflaton power spectrum}

The inflaton power spectrum $P_\phi(k,z)$ is defined by
\begin{equation}
\langle\delta\phi({\bf k},z)\delta\phi({\bf k}',z)\rangle=
P_\phi(k,z)(2\pi)^3\delta({\bf k}+{\bf k}')
\end{equation}
Substituting the first order inflaton perturbation from (\ref{usol}) and the correlation function
(\ref{cf}) gives
\begin{equation}
P_\phi(k,z)=(2\Gamma_{\rm eff}T)\,k^{-3}\int_z^\infty dz'G(z,z')^2(z')^{2-4\nu}\label{zin}
\end{equation}
in the slow-roll approximation.

Integrals of this type are examined in appendix \ref{appb}. The large $z$ approximation corresponds
to length scales much smaller than the horizon size, and we can use this to examine the flat space
limit,
\begin{equation}
P_\phi(k,z)\sim(2\Gamma_{\rm eff}T)\,k^{-3}{z\over 4(\nu-1)}\label{plz1}
\end{equation}
If the equivalence principal applies, then the result must agree with a calculation using flat space
thermal field theory (where the physical momentum is $ka^{-1}$),
\begin{equation}
P_\phi(k,z)=ka^{-1}Tk^{-3}.\label{plz2}
\end{equation}
This allows us to read off the value of the constant $\Gamma_{\rm eff}$ from a comparison of Eq.
(\ref{plz1}) and Eq. (\ref{plz2}) using Eq. (\ref{defnu}),
\begin{equation}
\Gamma_{\rm eff}=\Gamma+H.
\end{equation}
The value of the coefficient is required for any analysis of the density perturbations in the weak
regime($r<1$) of warm inflation. Such calculations have previously used the friction term
$\Gamma+3H$ for $\Gamma_{\rm eff}$, but fortunately this has little effect on the final power
spectrum \cite{Berera:2008ar}.

When $r$ is small and $\nu\approx 3/2$, the perturbation amplitude at horizon crossing ($z=1$) can
be found numerically,
\begin{equation}
P_\phi(k,1)\approx1.526\,k^{-3}\,HT
\end{equation}
When $r$ is large, the perturbation amplitude at horizon crossing is governed by a saddle
point in the integral (\ref{zin}), given according to eq. (\ref{saddle})
by,
\begin{equation}
k=a\,\sqrt{3\over 2}(H(t_F)\Gamma(t_F))^{1/2}.\label{freeze}
\end{equation}
We call the time in eq. (\ref{freeze}) the freezeout time $t_F$ for
the mode $k$. The freezeout time always precedes the horizon crossing time.
An analytic approximation to the integral valid for large values of $r$ is given in
eq. (\ref{fapprox}). With this we recover the result derived in \cite{Hall:2003zp},
improving on an earlier results in \cite{berera00},
\begin{equation}
P_\phi(k,1)\approx k^{-3}\,{\sqrt{\pi}\over2}(\Gamma H)^{1/2}T.
\end{equation}
An approximation which works remarkably well for both large and small values of $r$ is given by
\cite{Berera:2008ar}
\begin{equation}
P_\phi(k,1)\approx k^{-3}\,{\sqrt{\pi}\over2}H^{1/2}(\Gamma+3H)^{1/2}T.\label{pphi}
\label{papprox}
\end{equation}
We shall compare the accuracy of this approximation to numerical results in Sect. V.

\section{Coupled fluctuations}

In this section we analyse the density perturbations in the case where the friction coefficient in
the inflaton equation depends on temperature, i.e.
\begin{equation}
\Gamma\equiv\Gamma(T,\phi).
\end{equation}
Temperature dependence in the friction coefficient is a feature of many models
\cite{Moss:2006gt}, and numerical calculations have shown that this can
lead to interesting effects on the density perturbations \cite{Hall:2003zp}.
The temperature dependence of the friction coefficient is parameterised by a coefficient $c$,
\begin{equation}
c={T\Gamma_T\over \Gamma}\label{defc}
\end{equation}
The consistency of warm inflation requires only that $c<4$, and values of $c=3$ and $c=-1$ arise in
calculations of the reheating term for different models. 

The non-zero value of $c$ leads to additional terms in the fluctuation equations through
\begin{equation}
\delta\Gamma=\Gamma_T\delta T+\Gamma_\phi\delta \phi.\label{dgam}
\end{equation}
The extra $\delta\phi$ terms depend on the slow-roll parameter $\beta$ and they are first 
order in the slow-roll approximation, but the $\delta T$
terms are leading order and couple the perturbations of the inflaton and the radiation.

\subsection{Radiation fluctuations}

There are two sources of fluctuations in the radiation. The first source is
purely statistical and caused by the
microscopic particle motions. These fluctuations can be described as photon number fluctuations
\cite{bererafang95}, thermodynamic or quantum field fluctuations \cite{Hall:2003zp}, but all
approaches agree on the power spectrum $P_s$ for $\delta\rho_r$ in the flat space limit,
\begin{equation}
P_s=4\rho_rTa^{-3}.
\end{equation}
An estimate of the relative size of these fluctuations can be made
by comparing the statistical fluctuations to the scalar field potential fluctuations
$V_\phi\delta\phi$ calculated using Eq. (\ref{pphi}) at horizon crossing ($k=aH$),
\begin{equation}
{P_s\over V_\phi^2P_\phi}\sim{\rho_rTa^{-3} \over  V_\phi^2H^{1/2}(\Gamma+3H)^{1/2}Tk^{-3}}
\sim {r\over (1+r)^{3/2}}.
\end{equation}
These stochastic radiation fluctuations are relatively unimportant in both the weak ($r\ll 1$) and
strong ($r\gg 1$) warm inflationary regimes.

The second source of fluctuations, which turns out to play the dominant role, 
is due to the inhomogeneous energy flux from the inflaton field. 
We can describe these fluctuations using first order perturbation
theory. This can be formalated in many equivalent ways \cite{DeOliveira:2001he,Hall:2003zp}, 
but the formulation we use here is based on Ref. \cite{hwang02}, and Ref. 
\cite{Moss:2007cv}. 
The transfer of momentum and energy into the radiation is described by a
momentum flux $J$
and an energy flux $\delta Q$, given by
\begin{eqnarray}
J&=&-\Gamma\dot\phi\,\delta\phi,\label{defj}\\
\delta Q&=&\dot\phi^2\delta\Gamma\label{defdq}
\end{eqnarray}
If we keep only the leading terms in the slow roll
approximation, the metric perturbations drop out and
the energy density perturbations in a uniform
expansion gauge satisfy the equation
\begin{equation}
\delta\ddot\rho_{r}+9H\delta\dot\rho_{r}+
\left(20H^2+{1\over 3}k^2a^{-2}\right)\delta\rho_{r}=
k^2a^{-2}J+5H\delta Q+\delta \dot Q\label{drhor}
\end{equation} 
The energy flux (\ref{defdq}) can be expressed in terms of $\delta\rho_{r}$ using 
Eqs. (\ref{defc}) and (\ref{dgam})  by 
\begin{equation}
\delta Q=cH\delta\rho_r,
\end{equation}
The radiation fluctuations therefore satisfy
\begin{equation}
\delta\ddot\rho_{r}+(9-c)H\delta\dot\rho_{r}+
\left((20-5c)H^2+{1\over 3}k^2a^{-2}\right)\delta\rho_{r}=
- k^2a^{-2}(\Gamma\dot\phi)\,\delta\phi\label{drho}
\end{equation}
The equation for the inflaton fluctuations can simply be obtained by adding a term
$\delta\Gamma\dot\phi$ to Eq. (\ref{langa}),
\begin{equation}
\delta\ddot\phi+3H(1+r)\delta\dot\phi+
k^2a^{-2}\delta\phi+3cr(\Gamma\dot\phi)^{-1}\delta\rho_r=
(2\Gamma_{\rm eff} T)^{1/2}\xi\label{dphi}
\end{equation}
There are now two coupled equations which we can use to determine the fluctuation power
spectrum $P_\phi$.

An important feature of the coupled system is the growing mode $\delta\phi\propto a^{3c}$ present
for times such that $k\gg aH\gg k/r$. This replaces the constant mode responsible for freezing out
the perturbation amplitude on sub-horizon scales for the uncoupled system. The concept of
`freezout' is a useful one nevertheless, as it defines the time when the noise terms leave their
imprint on the final fluctuation amplitude. We will retain the terminology, but with the
understanding that `freezout' now means approaching the growing mode.

\subsection{Asymptotic expansions}

We shall determine the fluctuation amplitude for large $r$ by using a matched asymptotic expansion
technique. To
begin with, we convert the fluctuation equations to the variable $z=k/(aH)$ and drop the terms which
are first order in the slow-roll approximation,
\begin{eqnarray}
\delta\phi^{\prime\prime}-(3r+2)z^{-1}\delta\phi'
+\delta\phi+crz^{-2}(\Gamma\dot\phi)^{-1}\delta\rho_r
&=&(2\Gamma_{\rm eff} T)^{1/2}\hat\xi\label{eq1}\\
\delta\rho_r^{\prime\prime}-(8-c)z^{-1}\delta\rho_r'
+(20-5c)z^{-2}\delta\rho_r+{1\over 3}\delta\rho_{r}
+(\Gamma\dot\phi)\delta\phi&=&0\label{eq2}
\end{eqnarray}
We suppose that $r$  is a large parameter and $cr=O(r)$.

\subsubsection{The range $z<<r^{1/2}$}

In the range $z<<r^{1/2}$, the stochastic source terms are relatively unimportant and the leading
order contribution to Eq. (\ref{eq1}) for large $r$ is simply
\begin{equation}
\delta\rho_r\approx c^{-1}(\Gamma\dot\phi)z{\partial \over \partial z}\delta\phi.
\end{equation}
Substitute this into Eq. (\ref{eq2}) to get
\begin{equation}
z{\partial\over \partial z}\left(z{\partial\over \partial z}-5\right)
\left(z{\partial\over \partial z}+c-4\right)\delta\phi+
\frac13z^2\left(z{\partial\over \partial z}+{3c\over 2}\right)\delta\phi=0
\end{equation}
The solutions to this equation are Meijer-$G$ functions,
(see \cite{erdelyi}),
\begin{equation}
G^{31}_{13}\left({z^2\over 12}\left| \matrix{1-3c/2\cr2-c/2&0&5/2\cr}\right.\right),
\qquad
G^{30}_{13}\left(e^{\pm i\pi}{z^2\over 12}\left| \matrix{1-3c/2\cr2-c/2&0&5/2\cr}\right.\right).
\end{equation}
We are interested in the solution which grows most rapidly with time. Since the time evolution
proceeds from large to small $z$, this is the solution which decays most rapidly with $z$.

The asymptotic expansion of the  Meijer-$G$ functions for large $z$ can be found using standard
techniques, in particular
\begin{equation}
G^{31}_{13}\left({z^2\over 12}\left| \matrix{1-3c/2\cr2-c/2&0&5/2\cr}\right.\right)\sim
\left({z^2\over 12}\right)^{-3c/2}\Gamma_R(3c/2)\Gamma_R(3c/2+5/2)
\Gamma_R(2+c)\left\{1-{9c(3c+5)(c+2)\over z^2}+\dots\right\},\label{aeg}
\end{equation}
where $\Gamma_R$ is the Gamma function. When $c$ is positive, this is growing mode $\propto a^{3c}$.
For the other solutions
\begin{equation}
G^{30}_{13}\left(e^{\pm i\pi}{z^2\over 12}\left| \matrix{1-3c/2\cr2-c/2&0&5/2\cr}\right.\right)
\sim\sqrt{\pi}\left({z^2\over 12}\right)^{(3+c)/2}
e^{\mp iz/\sqrt{3}\mp i\pi(c-1)/2}.\label{aeg2}
\end{equation}
These oscillatory solutions grow more rapidly with time than the power-law solutions when $c<-3/4$.
We set
\begin{equation}
\delta\phi=AG^{31}_{13}\left({z^2\over 12}\left| \matrix{1-3c/2\cr2-c/2&0&5/2\cr}\right.\right)
+{\cal R}\,
BG^{30}_{13}\left(e^{ i\pi}{z^2\over 12}\left| \matrix{1-3c/2\cr2-c/2&0&5/2\cr}\right.\right)
\label{lowz}
\end{equation}
where $A$ and $B$ are constants and ${\cal R}$ indicates taking the real part.

\subsubsection{The range $1<<z<<r$}

In the range $1<<z<<r$ we have to take into account the noise term. If $c>-3/4$, then $\delta\rho_r$
has power-law behaviour and the leading terms in Eq. (\ref{eq2}) give,
\begin{equation}
\delta\rho_r\approx-3(\Gamma\dot\phi)\delta\phi.\label{rhophi}
\end{equation}
Substitute into Eq. (\ref{eq1}),
\begin{equation}
\delta\phi^{\prime\prime}-(3r+2)z^{-1}\delta\phi'
+\delta\phi-9crz^{-2}\delta\phi
=(2\Gamma T)^{1/2}\hat\xi
\end{equation}
As before, we solve this using a Green function,
\begin{equation}
G^c(z,z')={\pi\over 2}z^{\nu} z^{\prime\nu}
\left(J_{\nu+3c}(z)Y_{\nu+3c}(z')-J_{\nu+3c}(z')Y_{\nu+3c}(z)\right)
\quad\hbox{ for }z<z'.\label{gc}
\end{equation}
where $\nu=3(1+r)/2$. The power spectrum is given by
\begin{equation}
P^{c\ne 0}_\phi(k,z)=(2\Gamma T)k^{-3}\int_z^\infty dz'\,G^c(z,z')^2\,z^{\prime\, 2-4p}
\end{equation}
If we further restrict to $1<<z<<r^{1/2}$, then the integral can be approximated (see appendix A).
Comparing to the previous result for $c=0$, we get
\begin{equation}
{P^{c\ne 0}_\phi(k,z)\over P^{c= 0}_\phi(k,1)}\approx
{\Gamma_R(3c+3/2)\over\Gamma_R(3/2)}r^{3c}\left({3\over z^2}\right)^{3c}.\label{hiz}
\end{equation}
This result is not valid at horizon crossing ($z=1$), and so a matching procedure is required. When
$c<-3/4$ the solutions oscillate and Eq. (\ref{rhophi}) is no longer valid. The best we can do with
a simple analysis is to conclude from Eq. (\ref{aeg2}) that 
$P^{c\ne 0}_\phi(k,z)\propto (z^2/r)^{3+c}$, multiplied by the ensemble average of oscillatory
terms. We shall investigate this case numerically in the next section.

\subsubsection{Matching in the range $1<<z<<r^{1/2}$}

For $c>-3/4$, both approximations are valid (and agree) in the the range $1<<z<<r^{1/2}$. Comparing 
$\delta\phi^2$ from Eq. (\ref{lowz}) and $P_\phi$ from (\ref{hiz}) gives $B=0$ and
\begin{equation}
A^2=\left\{2^{3c} \Gamma_R(3c/2)\Gamma_R(3c/2+5/2)\Gamma_R(2+c)\right\}^{-2}
{\Gamma_R(3c+3/2)\over\Gamma_R(3/2)}\,r^{3c}\,P^{c= 0}_\phi(k,1).
\end{equation}
We are able to use Eq.  (\ref{lowz}) at $z=1$,
\begin{equation}
{P^{c\ne 0}_\phi(k,1)\over P^{c= 0}_\phi(k,1)}\approx
\left\{   { G^{31}_{13}\left({1\over 12}\left| \matrix{1-3c/2\cr2-c/2&0&5/2\cr}\right.\right)
\over 2^{3c} \Gamma_R(3c/2)\Gamma_R(3c/2+5/2)
\Gamma_R(2+c) } \right\}^2{\Gamma_R(3c+3/2)\over\Gamma_R(3/2)}\,r^{3c}.
\end{equation}
Numerical values of the result for different values of $c$ lie close to a simple power-law
approximation
\begin{equation}
{P^{c\ne 0}_\phi(k,1)\over P^{c= 0}_\phi(k,1)}\approx
\left({r\over r_c}\right)^{3c},\label{pc}
\end{equation}
where $r_c$ varies slowly in the range $0\le c\le3$, for example $r_1\approx8.53$, 
$r_2\approx7.66$ and   $r_3\approx 7.27$.

\subsection{Damping effects}

Damping effects can suppress the radiation fluctuations and may be important when the thermal
equilibrium of the radiation is imperfect. In warm
inflation thermal equilibrium is disrupted by particle production from the evolving inflaton field
which competes with the thermalising effects of self-interactions within the radiation gas
\cite{Graham:2008vu}. The effectiveness of the thermalisation process can be parameterised by the
value of
$H\tau_r$, where $\tau_r$ is the relaxation time of the radiation.

The most important damping effect in a pure radiation gas comes from the shear viscosity $\eta$
\cite{weinberg72}, which modifies the perturbation equation (\ref{drhor}) to,
\begin{equation}
\delta\ddot\rho_{r}+(9H+\eta k^2a^{-2})\delta\dot\rho_{r}+
\left(20H^2+{1\over 3}k^2a^{-2}\right)\delta\rho_{r}=
k^2a^{-2}J+5H\delta Q+\delta \dot Q,
\end{equation}
where \cite{weinberg72}
\begin{equation}
\eta={4\over 15}\rho_r\tau_r.
\end{equation}
The power-law behaviour of the growing mode is affected by the new damping term, so that a similar
analysis to the undamped case now gives
\begin{equation}
{P^{c\ne 0}_\phi(k,1)\over P^{c= 0}_\phi(k,1)}\approx
\left({r\over r_c}\right)^{3c'},
\end{equation}
where
\begin{equation}
c'={2c\over 1+(1+24cH\tau_r/5)^{1/2}}.
\end{equation}
The equilibrium is maintained when $H\tau_r\ll 1$ and in this case $c'\approx c$. However, the value
of $c'$ is still significant even when $H\tau_r\sim 1$ and the thermal equilibrium is very marginal.
Enhancement of the fluctuations therefore appears to be a fairly robust phenomenon.   

\section{Numerical simulation}

\begin{center}
\begin{figure}[htb]
\scalebox{0.45}{\includegraphics{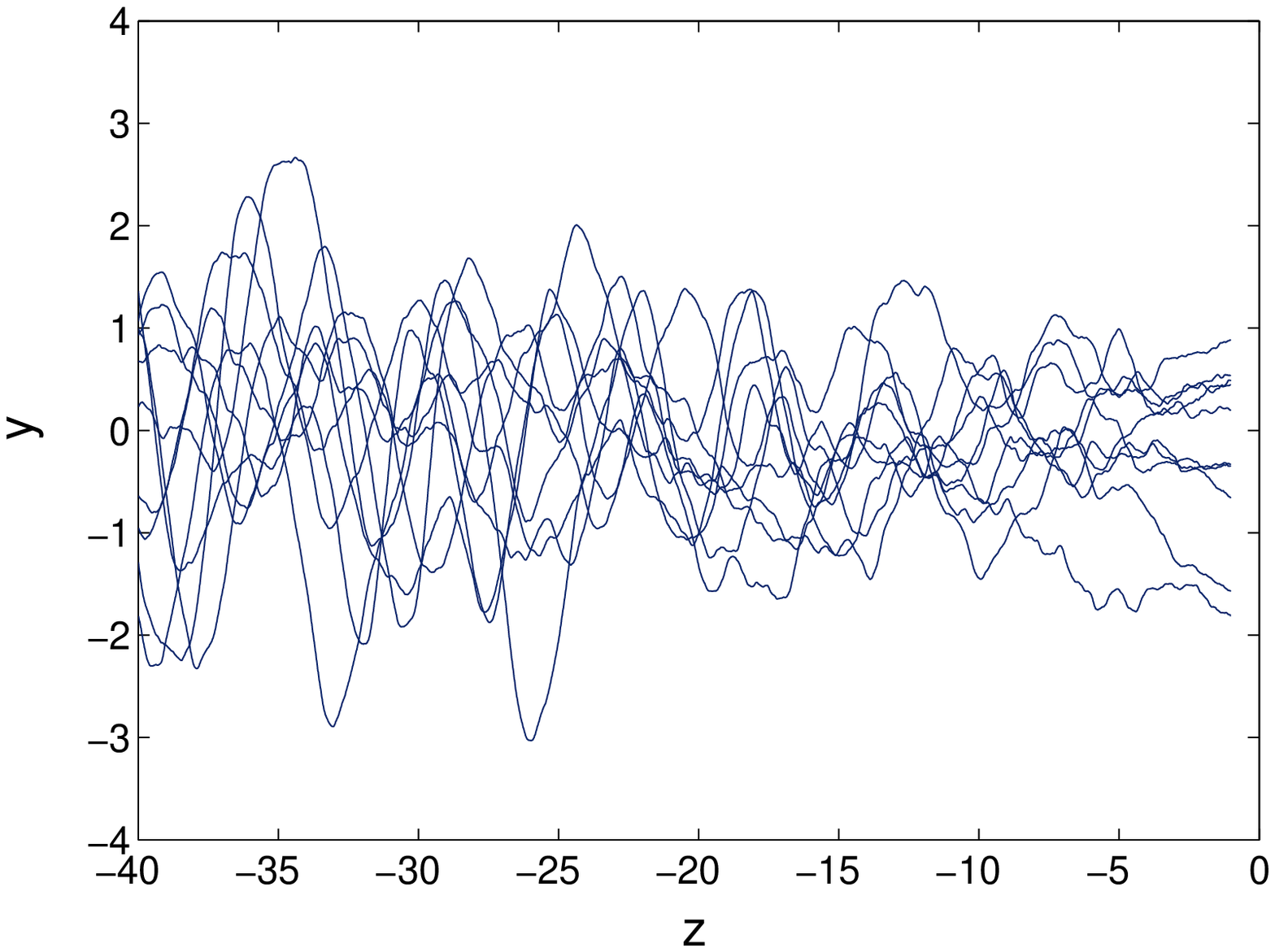}}
\scalebox{0.45}{\includegraphics{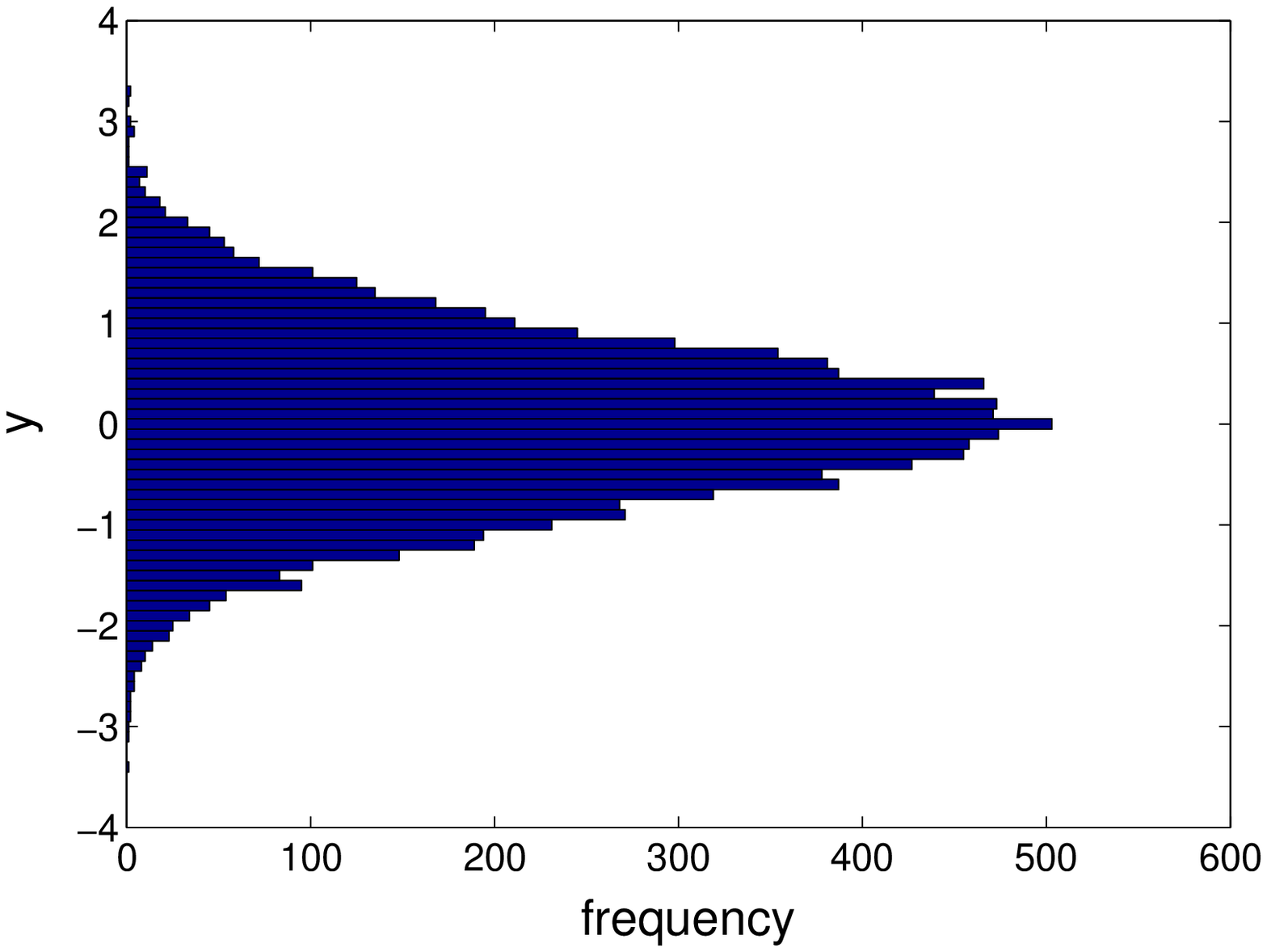}}
\caption{The time evolution of the inflaton perturbation for several runs with $c=1$ and $r=5$ is
shown on the left. The histogram on the right shows the values of the inflaton perturbation at
horizon crossing ($z=-1$) for $10000$ runs.}
\label{fig1}
\end{figure}
\end{center}

The full set of perturbation equations for warm inflation, including the metric perturbations, where
solved numerically by Hall et al. in ref. \cite{Hall:2003zp}. In the case where the parameter
$c=-1$, the power-spectrum of the perturbations had a series of oscillations superimposed on the
usual power-law behaviour in $k$.      

The computer code used in ref. \cite{Hall:2003zp} uses the stochastic
equations to set up the initial conditions and then evolves the full perturbation equations
deterministically. The drawback with this approach is that the code only runs once for each mode,
and does not necessarily produce the same results as an ensemble average of the stochastic system.
In this section we shall solve stochastic equations for the perturbations, but we work to leading
order in the slow-roll approximation and neglect the metric perturbations. The deterministic code
helps provide solid evidence that the
metric perturbations are small on sub-horizon scales, confirming the same conclusion obtained
analytically in ref. \cite{Moss:2007cv}. 

We shall use rescaled variables
\begin{equation}
y=k^{3/2}(2\Gamma_{\rm eff}T)^{-1/2}\delta\phi,\qquad 
w=k^{3/2}(2\Gamma_{\rm eff}T)^{-1/2}(\Gamma\dot\phi)^{-1}\delta\rho_r.
\end{equation}
The perturbation equations Eqs. (\ref{eq1}) and (\ref{eq2}) become
\begin{eqnarray}
y^{\prime\prime}-(3r+2)z^{-1}y'
+y+crz^{-2}w
&=&\hat\xi\\
w^{\prime\prime}-(8-c)z^{-1}w'
+(20-5c)z^{-2}w+{1\over 3}w
+y&=&0
\end{eqnarray}
where $\hat\xi$ is a normalised Gaussian random variable. The scalar field power spectrum is given
by
\begin{equation}
P_\phi=(2\Gamma_{\rm eff}T)\,k^{-3}\,\langle y(z)^2 \rangle
\end{equation}
Departures from a scale-free spectrum arise from the dependence of $\Gamma$, $H$ and $T$ on
the $k$-dependent horizon crossing (or freezeout) time. This can be determined from a numerical
solution of the background homogeneous field equations. Our aim here is to examine how
$\langle y(z)^2 \rangle$ depends on $r$ and $c$. 

\begin{center}
\begin{figure}[htb]
\scalebox{0.5}{\includegraphics{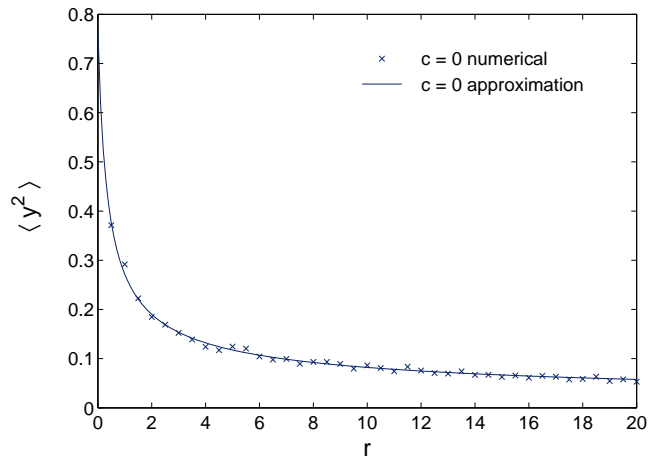}}
\caption{The variance of the inflaton at horizon exit from the numerical simulations has been
plotted against $r=\Gamma/3H$ for $c=0$ ($\Gamma$ independent of $T$). The data are averages over
$1000$ runs. The curve shows the approximation Eq. (\ref{yapprox}).}
\label{fig4}
\end{figure}
\end{center}

\begin{center}
\begin{figure}[htb]
\scalebox{0.5}{\includegraphics{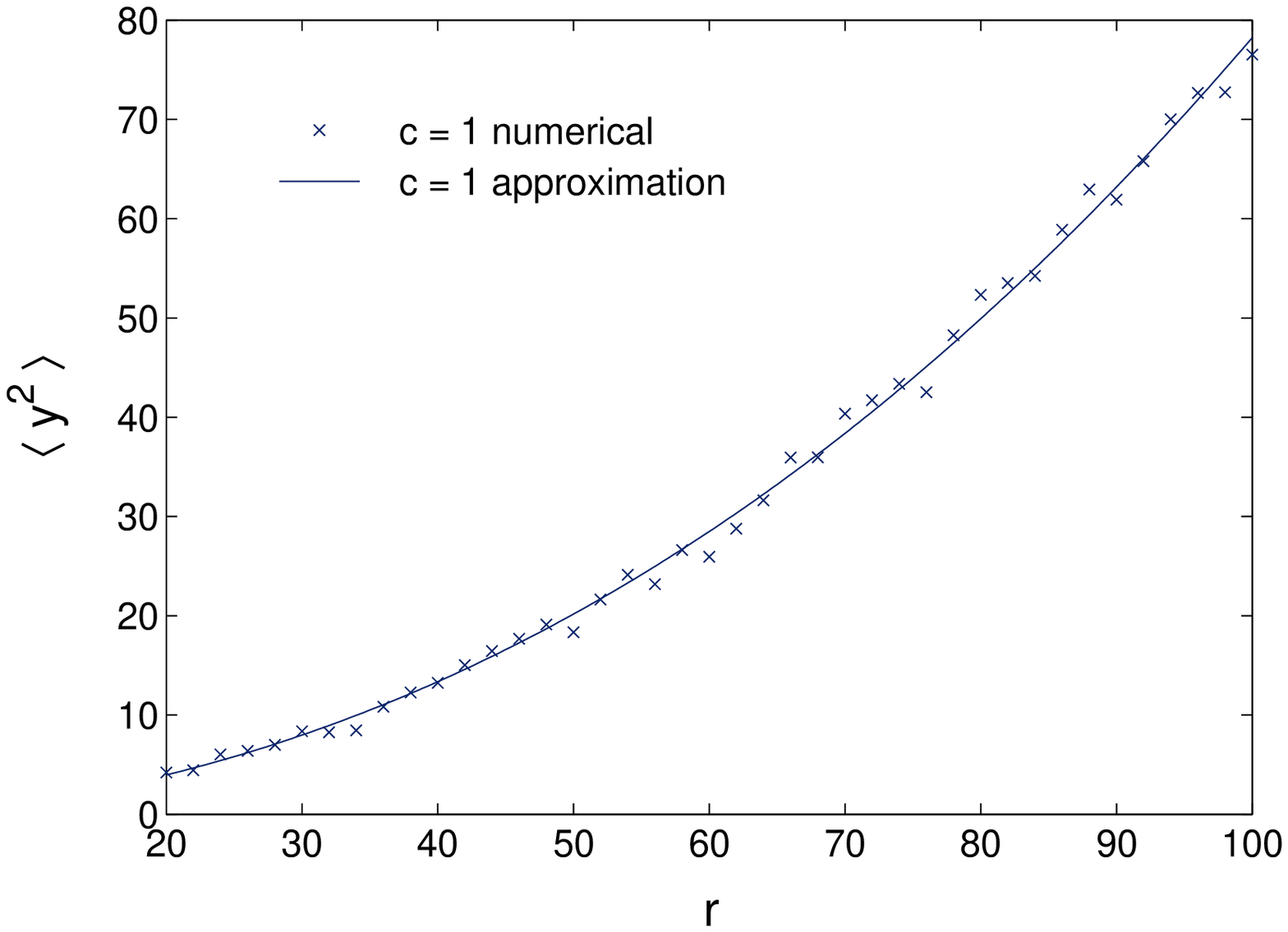}}
\caption{The variance of the inflaton at horizon exit from the numerical simulations has been
plotted against $r=\Gamma/3H$ for $c=1$ ($\Gamma\propto T$).
The data are averages over $1000$ runs. The curve shows the numerical fit
$A_1r^{5/2}+B_1r^{3/2}$.}
\label{fig2}
\end{figure}
\end{center}

\begin{center}
\begin{figure}[htb]
\scalebox{0.5}{\includegraphics{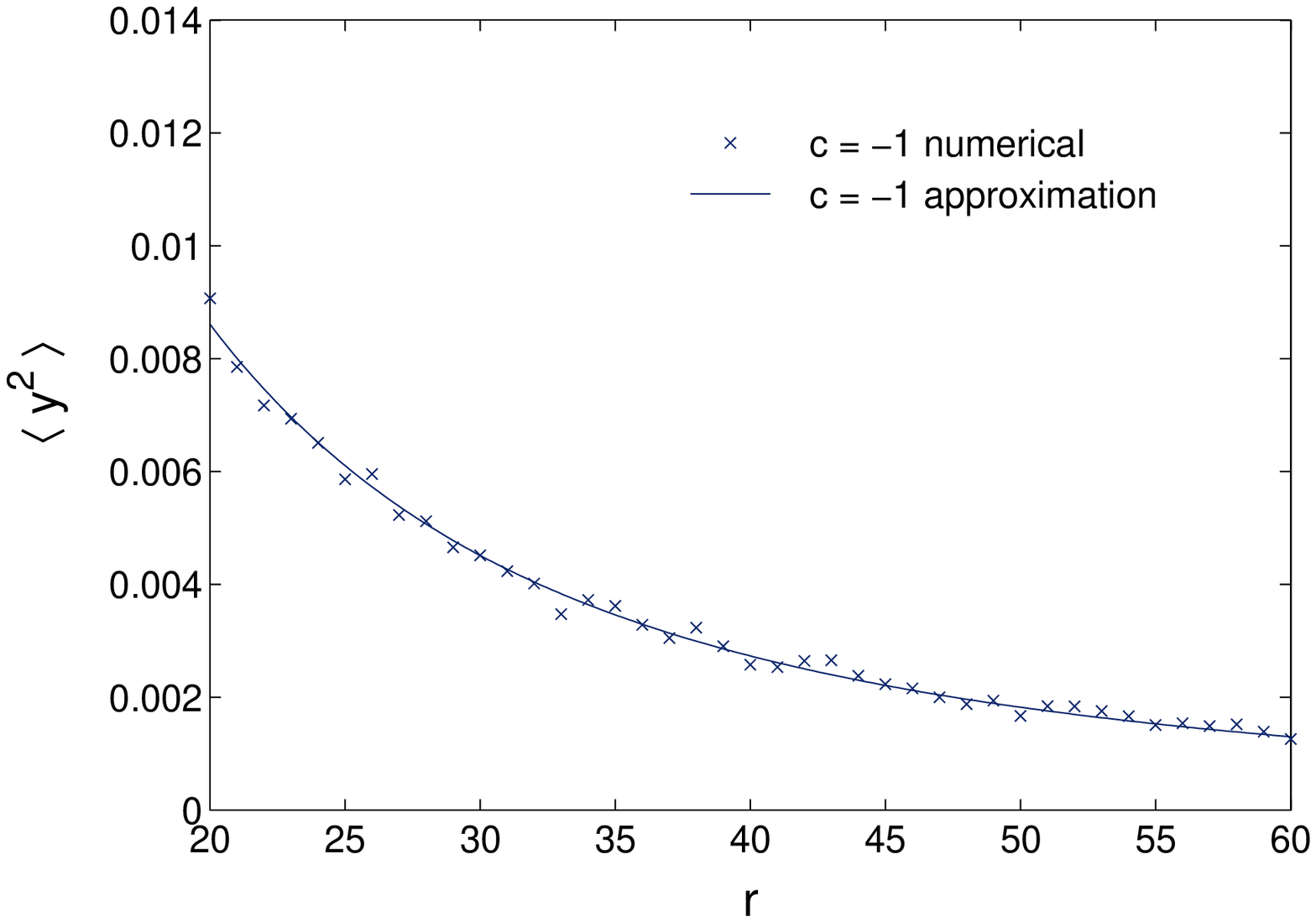}}
\caption{The variance of the inflaton at horizon exit from the numerical simulations has been
plotted against $r=\Gamma/3H$ for  $c=-1$ ($\Gamma\propto T^{-1}$).
The data are averages $1000$ runs. The curve shows the numerical fit 
$A_{-1}r^{-2}+B_{-1}r^{-3}$.}
\label{fig6}
\end{figure}
\end{center}

Numerical results for $c=0$, $c=1$ and $c=-1$ are shown in figures \ref{fig1}-\ref{fig6}. These
have been obtained using a fourth order Runge-Kutta integration scheme and the Box-Muller algorithm
for generating Gaussian random variables. Fig. \ref{fig1} shows the time evolution of the inflaton
field, plotted against $z=-k/(aH)$ for convenience. The histogram on the right shows the
distribution of field values at horizon crossing $z=-1$ and fits a Gaussian distribution. The other
graphs show $\langle y^2\rangle$
at horizon crossing for different values of $r=\Gamma/3H$.

The curves superimposed on the data are based on the theoretical results  obtained in the
previous section.  For $c> -3/4$, Eq. (\ref{pc}) when converted into the rescaled variables
corresponds to
\begin{equation}
\langle y^2 \rangle\approx\frac14\left({\pi\over 3}\right)^{1/2}r_c^{-3c}r^{3c-1/2}.
\label{ysapprox}
\end{equation}
The theoretical prediction is excellent for $c=0$ and $r\ll 1$. The data
can also be approximated by rational fractions over the entire range, for example the approximation
\begin{equation}
\langle y^2 \rangle\approx{\sqrt{3\pi}\over 4}{(1+r)\over 1+3r}^{1/2}\label{yapprox}
\end{equation}
works extremely well for $c=0$ and leads to the prediction for the inflaton power spectrum given in
Eq.  (\ref{pphi}).

The data for the $c=1$ case have been fit to a curve of the form $A_1r^{5/2}+B_1r^{3/2}$ based on
the theoretical prediction. The need for next-to-leading order corrections to the asymptotic
formula is indicated by the analysis, for
example in Eq. (\ref{aeg}). The coefficients for the best fit are shown in table \ref{table1}, and
correspond to $r_c=8.2$ in Eq. (\ref{ysapprox}), which is $5\%$ from the theoretical prediction
$r_c=8.5$. Results for $c=2$ and $c=3$ using the function
$A_cr^{3c-1/2}+B_cr^{3c-3/2}$ are also shown in the table.

The most significant feature in the data for the $c=-1$ case is that we have found no evidence for
the oscillations in the amplitude found previously in Ref. \cite{Hall:2003zp}. This is presumably
due to the fact that the numerical code used in Ref. \cite{Hall:2003zp} only performs one run per
mode, whereas we average over many runs.
The results have been fit to a curve of the form $A_{-1}r^{-2}+B_{-1}r^{-3}$, in accordance with the
leading order $(z^2/r)^{3+c}$ behaviour of $\delta\phi^2$ suggested in the previous section. The
coefficients for the best fit are given in table \ref{table1}.

\begin{table}[htb]
\caption{\label{table1}Coefficients in the numerical fits for the variance of the inflaton at
horizon exit with their range of validity. }
\begin{ruledtabular}
\begin{tabular}{llll}
$c$&$A_c$&$B_c$&range\\
\hline
$3$&$4.51\times 10^{-9}$&$3.34\times10^{-6}$&$r>100$\\
$2$&$1.22\times 10^{-6}$&$4.39\times10^{-4}$&$r>20$\\
$1$&$4.66\times 10^{-4}$&$2.94\times10^{-2}$&$r>10$\\
$-1$&$5.30$&$-37.1$&$r>20$\\
\end{tabular}
\end{ruledtabular}
\end{table}

\section{Density fluctuations}

We have seen that the small scale inflaton fluctuations freeze out by the time they cross the
horizon. Whilst these fluctuations are freezing out, the
metric fluctuations are relatively small (in the uniform expansion-rate gauge, for example). 
Eventually, the wavelength of the perturbations becomes far larger than the effective
cosmological horizon and the metric perturbations become important. On these large scales
it becomes possible to use a small-spatial-gradient expansion (first
formalised by Salopeck and Bond \cite{Salopek:1990jq}). This approach allows
us to define the curvature perturbation $\zeta$ so that it is conserved even
in the non-linear theory 
\cite{Sasaki:1995aw,Lyth:2004gb,Lyth:2005fi}. 
The spectrum can be obtained, to high
accuracy, by matching the small and large scale approximations at horizon
crossing. 

The fluctuations can be described entirely by the conserved
expansion fluctuation $\zeta$ on constant density hypersurfaces \cite{Bardeen:1980kt}, which is
defined for general hypersurfaces by
\begin{equation}
\zeta=\frac12\ln\left(1+2\varphi\right)+\frac13\int{d\rho\over p+\rho},
\end{equation}
where $\varphi$ is the spatial curvature perturbation. After 
using the slow roll equations,
\begin{equation}
\zeta=\frac12\ln\left(1+2\varphi\right)+\int {H\over\dot\phi}d\phi,
\end{equation}
where $H\equiv H(\phi)$ and $\dot\phi\equiv \dot\phi(\phi)$ are given from eqs.
(\ref{sr1}-\ref{sr3}).

Consider a uniform curvature gauge $\varphi=0$ and
$\zeta\equiv\zeta(\phi)$. When the inflaton perturbations are expanded as
before in eq. (\ref{ipe}), we have 
\begin{equation}
\zeta=\zeta_\phi\delta\phi
\end{equation}
where $\phi$ subscripts denote derivatives with respect to $\phi$ and 
$\zeta_\phi=H/\dot\phi$. Hence the power spectrum of density
perturbations is
\begin{equation}
P_\zeta(k)=\zeta_\phi^2 P_\phi(k)
\end{equation}
In the previous section we evaluated the small scale inflaton perturbations
in a uniform expansion-rate gauge, but on sub-horizon scales the metric perturbations 
are relatively small and the inflaton perturbations in the uniform expansion-rate gauge and
the uniform curvature gauge are approximately equal.

In the strong regime of warm inflation we can use Eq (\ref{pc}) for the inflaton power spectrum when
$r>r_c$ and $c>-3/4$, to give
\begin{equation}
P_\zeta(k)={\sqrt{3\pi}\over2}{H^3T\over \dot\phi^2}\left({r\over r_c}\right)^{3c}r^{1/2}\,k^{-3},
\end{equation}
This can be inverted, using the slow-roll equations (\ref{sr1}-\ref{sr3}),
to express the basic energy scales of the inflationary model in terms of the fluctuation amplitude,
$r$ and slow-roll parameters. The value of the potential when the mode with wave number  $k$
crosses the horizon is given by 
\begin{equation}
V^{1/4}=g_*^{1/12}\epsilon^{1/4}r^{-3/4}\left({r\over r_c}\right)^{-c}
\left({P_\zeta(k)k^3\over 10^{-9}}\right)^{1/3}
(3.73\times 10^{15}{\rm GeV}),
\end{equation} 
for $r>r_c$ and $c>-3/4$.   The $c$ dependence depresses the energy scale of inflation relative to
models with $c=0$. In most models this leads to an increase in $\Gamma/3H$ and $T/H$, making warm
inflation stronger than in the $c=0$ case. When $c<0$ the energy scale of inflation increases and
the warm inflation becomes weaker. On the other hand, if primordial tensor perturbations are seen
in the cosmic microwave background, this would favour a large energy scale for inflation and $c<0$.

\section{conclusion}

We have analysed the thermal fluctuations present in warm inflationary scenarios to obtain results
for the primordial power spectrum using both analytical techniques and numerical simulations. For
the first time it has been possible to present general formulae for the power spectrum in models
where the fiction coefficient in the inflaton equation of motion depends on temperature. These
results are crucial for building models of warm inflation which are consistent with cosmic
microwave background observations.

The results for the primordial power spectrum of curvature fluctuations are as follows. For a
friction coefficient $\Gamma$ which is independent of temperature, the familiar approximation
\begin{equation}
P_\zeta(k)={\sqrt{\pi}\over2}{H^{5/2}(\Gamma+3H)^{1/2}T\over \dot\phi^2}\,k^{-3},\label{ops}
\end{equation} 
fits both analytic and numerical calculations remarkably well for all values of $\Gamma$. This
result can be applied whenever thermal fluctuations are the dominant source of curvature
fluctuations, which is typically when $T>H$.

The power spectrum is given by a different formula when the friction coefficient depends on
temperature. For $\Gamma\propto T^c$, and $r=\Gamma/3H\gg1$, the amplitude is enhanced when $c>0$
and reduced when $c<0$. For accurate results we recommend using
\begin{equation}
P_\zeta(k)=6{H^3T\over \dot\phi^2}f(r)\,k^{-3},
\end{equation}
where the function $f(r)\sim A_cr^{3c+1/2}+B_cr^{3c-1/2}$ for $c>0$  and the coefficients are given
in table \ref{table1}. 

For less accurate work, a reasonable approximation for $c>0$ is provided by
\begin{equation}
P_\zeta(k)={\sqrt{3\pi}\over2}{H^3T\over \dot\phi^2}(1+r)^{1/2}\left(1+{r\over r_c}\right)^{3c}
\,k^{-3},
\end{equation}
with $r_3\approx7.3$. This formula is consistent  with the old result given by Eq. (\ref{ops}) for
$c=0$ and approaches the numerical results for large $r$.

The most promising models of warm inflation to date use the `two-stage' decay mechanism, for which
the friction coefficient has $c=3$ \cite{BasteroGil:2006vr,BasteroGil:2009ec}. The new power
spectrum is very different from the old result Eq. (\ref{ops}), and therefore these models must be
analysed using the new power spectrum. The effect of the new formula is to reduce the inflationary
scale, and it is likely that the restrictions on warm inflationary models will be relaxed.

It is known that warm inflationary models produce a significant amount of non-gaussianity, possibly
enough to be observed in microwave background experiments \cite{Moss:2007cv,Moss:2007qd}. So far, the analysis
of non-gaussianity has only been done  in cases where the friction coefficient is independent of
temperature. However, it seems possible that the analysis of the coupled perturbation equations
initiated here could be extended to second order to obtain results for the non-gaussianity when the
friction coefficient is temperature dependent.

\appendix

\section{Integrals}\label{appb}

We begin with an approximation to the integral
\begin{equation}
F(z)=
\int_{z}^\infty dz'
G(z,z')^2
(z')^{2-4\nu}
\end{equation}
where $\nu=3(1+r)/2$ and the retarded Green function $G$ is given in eq.
(\ref{rgf}). The leading terms for large $\nu$ and fixed $z$,
come from
\begin{equation}
F(z)\approx
{\pi^2\over 4}
z^{2\nu} Y_{\nu}(z)^2
\int_0^\infty J_{\nu}(z')^2z^{\prime 2-2\nu}dz'
\end{equation}
This is a standard integral,
\begin{equation}
\int_0^\infty J_{\nu}(z')^2
z^{\prime 2-2\nu}dz'=
{\Gamma_R(\nu-1)
\over 4\Gamma_R(2\nu-1/2)\Gamma_R(\nu-1/2)}
\end{equation}
where $\Gamma_R$ is the gamma function. We also have
\begin{equation}
z^\nu Y_\nu(z)\sim 
-{2^\nu\over\pi\Gamma_R(\nu)}\left(1+{z^2\over 4\nu}+\dots\right).
\end{equation}
Hence, 
\begin{equation}
F(z)\sim
\sqrt{\pi\over 32\nu}
\left(1+{z^2\over 2\nu}+\dots\right).\label{fapprox}
\end{equation}
A similar integral arises when we replace $G$ with the Green function $G^c$ in (\ref{gc}), and in
this the case the integral, which we now call $F^c(z)$, has an asymptotic expansion with leading
term
\begin{equation}
F^c(z)\sim
{\Gamma_R(c+3/2)\over\Gamma_R(3/2)}\left({2\nu\over z^2}\right)^{3c}\sqrt{\pi\over 32\nu}.
\end{equation}
in the range $z\ll \nu^{1/2}$.

A slightly more challenging problem is the integral
\begin{equation}
F(z,g)=
\int_z^\infty 
G(z,z')^2
z^{\prime\,2-4\nu}g(z')\,dz'
\end{equation}
where $g(z)$ is a smooth function. We proceed as above to get 
\begin{equation}
F(z,g)\approx
{\pi^2\over 4}z^{2\nu}Y_{\nu}(z)^2
\int_0^\infty J_{\nu}(z')^2
z^{\prime 2-2\nu}g(z')dz'.
\end{equation}
For large $\nu$ there is a Debye approximation for the Bessel functions which
is valid in the range $0<z<\nu$, 
\begin{equation}
J_\nu(z)\sim (2\pi\nu\tanh\alpha)^{-1/2}e^{\nu(\tanh\alpha-\alpha)}
\end{equation}
where $\nu\,{\rm sech}\alpha=z'$. The relevant part of the
integral becomes
\begin{equation}
F(z,g)\approx
{\pi\over 8}z^{2\nu}Y_{\nu}(z)^2
\nu^{2-2\nu}\int_0^\infty(\cosh\alpha)^{2\nu-3}
\,e^{2\nu(\tanh\alpha-\alpha)}\,g(z')\,d\alpha
\end{equation}
where $\nu\,{\rm sech}\alpha=z'$. We find that there is a saddle
point in this range at the value of $\alpha$ corresponding to $z=z_F$,
where
\begin{equation}
z_F=(3\nu)^{1/2}\label{saddle}
\end{equation}
Expanding about the saddle point gives
\begin{equation}
F(z,g)\sim
\sqrt{\pi\over 32\nu}g(\tau_F)
\end{equation}
which agrees with our earlier result when $g(z)=1$. This saddle point
is responsible for the phenomenon of `freezing out' of the thermal
fluctuations. The value of $z$ decreases with time and the fluctuations
always freeze out before they cross the horizon at $z=1$.

\bibliography{paper.bib,cosper.bib}

\end{document}